\documentstyle{proceeding}
\input psfig.tex

\begin{document}
\title{\centerline{TWO DISTINCT MODES IN THE LOW (HARD)}
\centerline{STATE OF}
\centerline{CYGNUS~X-1 AND 1E1740.7-2942}}
\author{S.~Kuznetsov\inst{1,2},~M.~Gilfanov\inst{1},~E.~Churazov\inst{1},~R.~Sunyaev\inst{1},~A.~Vikhlinin\inst{1,2},~N.~Khavenson\inst{1},
A.~Dyachkov\inst{1}, P.~Laurent\inst{3}, A.~Goldwurm\inst{3},
B.~Cordier\inst{3}, M.~Vargas\inst{3}, P.~Mandrou\inst{4},
J.P.~Roques\inst{4}, E.~Jourdain\inst{4}, V.~Borrel\inst{4}.}

\institute{Space Research Institute, Russian Academy of Sciences Profsoyuznaya
84/32, 117810 Moscow, Russia
\and Moscow Physical-Technical Institute, Institutskiy Lane 9,
Dolgoprudny, 141700 Moscow area, Russia
\and Service d'Astrophysique, DAPNIA/DSM, Bt 709, CEA Saclay,
91191 Gif sur Yvette Cedex, France
\and Centre d'Etude Spatiale des Rayonnements 9, avenu du Colonel Roche,
BP 4346, 31029 Toulouse Cedex, France}
\maketitle
%%%%%%%%%%%%%%%%%%%%%%%%%%%%% A B S T R A C T %%%%%%%%%%%%%%%%%%%%%%%%%%%%%%
\begin{abstract}

The entire dataset of the GRANAT/SIGMA observations of Cyg X-1 and
1E1740.7-2942 in 1990--1994 was analyzed in order to search for correlations
between primary observational characteristics of the hard X-ray (40--400
keV) emission -- hard X-ray luminosity $L_X$, hardness of the spectrum
(quantified in terms of the best-fit thermal bremsstrahlung temperature
$kT$) and the {\em rms} of short-term flux variations.

Two distinct modes of the $kT$ vs. $L_X$ dependence were found for both
sources (Fig.1). At low luminosity -- below the level corresponding
approximately to the $\gamma_1$ state of Cyg X-1 (Ling et al. 1987) -- the
$kT$ increases as the $L_X$ increases.  Quantitatively it corresponds to
increase of the temperature from 70 keV at $\approx 0.5L_{\gamma_1}$ to 150
keV at $\approx 1.2L_{\gamma_1}$.  Above the luminosity level of $\approx
1.2L_{\gamma_1}$ the spectrum hardness is nearly constant ($T\approx 150$
keV) and does not depend on the luminosity.

In the case of Cyg X-1 (1E1740.7-2942 is not bright enough and is located in
crowded Galactic Center region) the correlation of similar kind was found
between the spectrum hardness and $rms$ of the short-term flux variations
(Fig.2).  The increase of the $kT$, corresponding to the increasing branch
on the $kT$ vs. $L_X$ diagram, is accompanied with increase of the $rms$
from $\la$ few percent level to $\approx 10-15$\%. Further increase of the
$rms$ is not accompanied with change of the $kT$ and does not correlate with
changes in the luminosity.
\end{abstract}

\section{Observations}

\begin{figure}
\psfig{figure=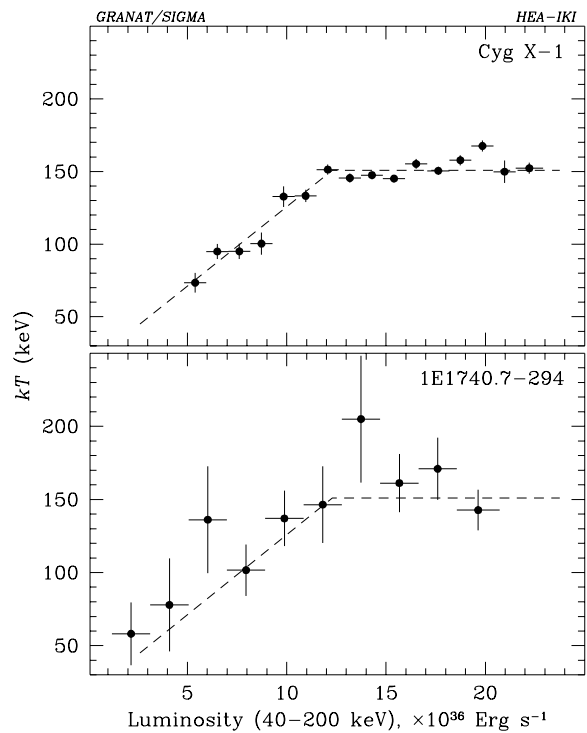,height=9.8truecm}
\caption[]{The hardness of the spectrum
versus hard X-ray luminosity (40-200 keV) for Cyg~X-1 (upper panel) and
1E1740-294 (lower panel).  The luminosity for each bin was calculated using
corresponding best fit thermal bremsstrahlung spectrum. The broken constant
best fit to Cyg~X-1 data is shown in both panels by the dashed lines.}
\end{figure}

In order to search for correlation between various characteristics of the
hard X-ray emission from the source the original data were regrouped
according to the source intensity in the following way. The entire range of
the 40-150 keV flux variations was divided into number of bins of the same
width. For each individual dataset the bin number was determined according
to its intensity. The mean energy and power density spectra
corresponding to each intensity bin  were calculated by averaging  over all
individual datasets with intensity falling into the given bin intensity
range. 

For Cyg X-1 16 intensity bins were chosen covering the 1.9 to 6.9
cnt/sec/cm$^{2}\times10^{-2}$\ (0.5-1.8 Crab) intensity range. The
regrouping procedure was applied to the data of individual SI exposures (4-8
hours long - the highest time resolution providing both spectral and timing
information) each exposure being treated as a separate dataset.

In the case of 1E1740-294, having 5 to 20 times lower signal to noise ratio
the data averaged over single observations (comprised of 1-6 SI exposures
with total duration of 4-34 hours) were treated as individual datasets to be
regrouped. The intensity range 0.3 to 5.6 cnt/sec/cm$^{2}\times10^{-3}$\
(80-1500 mCrab) was divided into 10 intensity bins.

\section{Spectral and timing analysis}

In order to quantify the hardness of the spectrum the optically-thin thermal
bremsstrahlung model was chosen. Although likely having no direct physical
relation to the origin of the hard X-ray emission from the source, it
provides a good approximation to the observed spectra and quantifies the
spectral shape in terms of a single parameter - the best-fit temperature.
The relative error of the observed spectrum approximation by this model is
less than 10\% in the 40-300 keV energy range. The approach to timing
analysis of the SIGMA data was described in Vikhlinin et al. (1994).

\section{Results}

The dependence of the best-fit bremsstrahlung temperature on the 40-200 keV
luminosity for both sources is shown in Fig.1. The similarity in the
behaviors of the two sources is apparent. Approximation of the points for
Cyg X-1 by a broken constant is shown by a dashed line. Surprisingly, the
same curve describes very well the data for 1E1740-294.

\begin{figure}
\psfig{figure=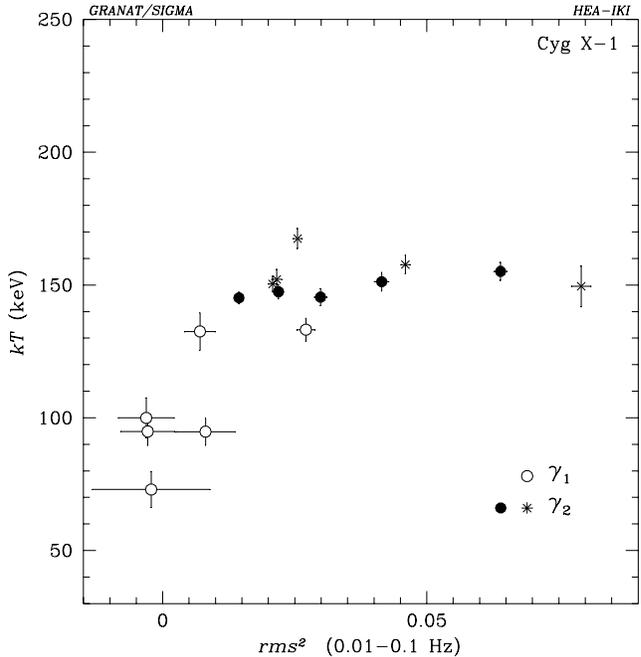,height=9.3truecm}
\caption[]{The best fit bremsstrahlung temperature plotted
against the $rms^{2}$ of the flux variation for Cyg~X-1.  The open circles
correspond to the source flux $\la$ the $\gamma_1$ state level,
the solid circles - between the $\gamma_1$~ and $\gamma_2$~ levels, the
starred - above the $\gamma_2$~ level.} 
\end{figure} 

The Fig. 2 shows a dependence of the best-fit bremsstrahlung temperature
upon the $rms$ of the short-term flux variations for Cyg X-1 (the individual
datasets were grouped in the same way as in Fig.1).

It should be noted, that shown in Fig. 1 and 2 are parameters derived from
averaged energy and power density spectra for each intensity bin and the
error bars are statistical only. The analysis of individual datasets,
corresponding to the given intensity bin reveals considerable dispersion of
the best-fit parameters above the level of statistical fluctuations. This
dispersion is of the order of $\sim$15\% of the values of the best-fit
temperature shown in Fig. 1. Therefore, dependencies, shown in Fig. 1 and 2
are not point-to-point correlations, but rather represent some averaged
pattern of the behavior of the parameters.

There are two major possibilities to explain observed behavior. the first
one is that softening of the source spectrum corresponds to the beginning of
the transition to the soft spectral state known as a common feature of black
hole candidates. It is known also that the $rms$ of the short-term flux
variations is considerably smaller during soft spectral state which explains
the pattern in Fig. 2. Note, that in this case the decrease of the hard
X-ray luminosity does not reflect at all the behavior of the overall X-ray
luminosity which is increasing.

The second possibility is that the change of the spectral hardness reflects
changes of conditions in the hard X-rays production site having no
connection with transition to the soft spectral state. In this case pattern
in Fig. 2 hints on close relation between the hardness of the emergent
spectrum and level of the short-term flux variations.

\begin{acknowledgements}
This work was supported in part by International Science Foundation grants
M4W000 and M7M000 and RBRF grant 93-02-17166.
\end{acknowledgements}

\end{document}